\documentclass[aip,apl,preprint]{revtex4-1}
\usepackage{graphicx}
\usepackage{natbib}
\usepackage{amsmath}
\usepackage{amssymb}
\usepackage{bm}
\usepackage{color}

\begin{document}

\title{Thermal Expansion Coefficients of Bi$_2$Se$_3$ and Sb$_2$Te$_3$ Crystals from 10 K to 270 K}

\author{X. Chen}
\email{xcchen@gatech.edu}
\affiliation{School of Physics, Georgia Institute of Technology, Atlanta, GA 30332}

\author{H. D. Zhou}
\affiliation{National High Magnetic Field Laboratory, Tallahassee, FL 32310}

\author{A. Kiswandhi}
\affiliation{National High Magnetic Field Laboratory, Tallahassee, FL 32310}

\author{I. Miotkowski}
\affiliation{Department of Physics, Purdue University, West Lafayette, IN 47907}

\author{Y. P. Chen}
\affiliation{Department of Physics, Purdue University, West Lafayette, IN 47907}

\author{P. A. Sharma}
\affiliation{Sandia National Laboratories, Albuquerque, NM 87185}

\author{A. L. Lima Sharma}
\affiliation{Sandia National Laboratories, Albuquerque, NM 87185}

\author{M. A. Hekmaty}
\affiliation{Sandia National Laboratories, Livermore, CA 94550}

\author{D. Smirnov}
\affiliation{National High Magnetic Field Laboratory, Tallahassee, FL 32310}

\author{Z. Jiang}
\email{zhigang.jiang@physics.gatech.edu}
\affiliation{School of Physics, Georgia Institute of Technology, Atlanta, GA 30332}


\begin{abstract}
Lattice constant of Bi$_2$Se$_3$ and Sb$_2$Te$_3$ crystals is determined by X-ray powder diffraction measurement in a wide temperature range. Linear thermal expansion coefficients ($\alpha$) of the crystals are extracted, and considerable anisotropy between $\alpha_\parallel$ and $\alpha_\perp$ is observed. The low temperature values of $\alpha$ can be fit well by the Debye model, while an anomalous behavior at above 150 K is evidenced and explained. Gr\"uneisen parameters of the materials are also estimated at room temperature. 
\end{abstract}

\maketitle

Recently, much attention has been given to an intriguing class of materials, the so-called topological insulators (TIs). This type of material exhibits a band gap in the bulk, but gapless states on the edge or surface, which are protected by topological order and cannot be analogized to conventional semiconductors or insulators \cite{Hasan_RMP_2010,Qi_arxiv}. Bi$_2$Se$_3$, Bi$_2$Te$_3$ and Sb$_2$Te$_3$ are among the most interested compounds of three-dimensional TIs, owing to their robust and simple surface states \cite{Zhang_natphys}. Although these compounds were under extensive studies in 1950s and 1960s as excellent thermoelectric materials, some basic physical properties still remain unexplored. In this letter, we present the measurements of the temperature dependent linear thermal expansion coefficients of Bi$_2$Se$_3$ and Sb$_2$Te$_3$ crystals using X-ray powder diffraction (XRD). Thermal expansion is the tendency of materials to change in size and shape as they heat and cool. It is essential to device design and engineering, as the induced strain could cause the deformation of the device and affect its phonon dynamics. Indeed, our recent Raman spectroscopy study of TIs has uncovered significant contributions in the temperature dependent phonon frequency shifts from the thermal expansion of the materials \cite{Raman_Kim}. In addition, the knowledge of thermal expansion coefficients is necessary for the directional growth of TI crystals and the understanding of the high thermoelectric efficiency \cite{pavlova}.

\begin{figure}[t]
\includegraphics{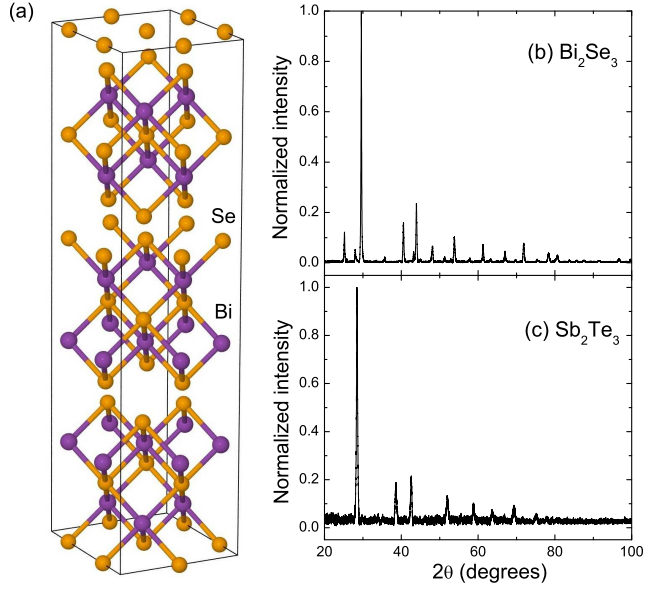}
\caption{\label{structure}(color online) (a) Hexagonal unit cell of Bi$_2$Se$_3$. The distance between two nearest Se atoms of the two adjacent Se layers is larger than the sum of their covalent radii. As a result, the Se-Bi-Se-Bi-Se quintuple layers are hold together by weak van der Waals force. Bi$_2$Te$_3$ and Sb$_2$Te$_3$ have a similar structure by substituting Se atoms with Te, Bi atoms with Sb. Right panel: XRD spectra of (b) Bi$_2$Se$_3$ and (c) Sb$_2$Te$_3$ at 10 K. Maximum peak intensity is normalized to 1.}
\end{figure}

Large grain polycrystalline Bi$_2$Se$_3$ materials (single crystal grain size $>$1 mm) were synthesized at Sandia National Laboratories. First, Bi$_2$Se$_3$ pieces (99.999\%, from VWR international, LLC.) were placed in an evacuated ($<$10$^{-7}$ Torr) quartz ampoule and melted at 800 $^\circ$C for 16 hours. The melt was then cooled at 10 $^\circ$C/h to 550 $^\circ$C, held for 3 days at this temperature, and finally allowed to furnace cool to room temperature. Single crystals of Sb$_2$Te$_3$ were grown by Bridgman method at Purdue University. Stoichiometric amount of high purity antimony and tellurium (99.999\%) was deoxidized and purified by multiple vacuum distillations under dynamic vacuum of $<$10$^{-7}$ Torr, and then heated up to 900 $^\circ$C. This was followed by a slow cool down under a controlled pressure to minimize tellurium defects. Afterwards, the crystals were grown at a speed of 0.5-1.5 mm/h with a linear temperature gradient set to 5 $^\circ$C/cm. Bi$_2$Se$_3$ and Sb$_2$Te$_3$ crystals have similar rhombohedral structure with five atoms in the trigonal primitive cell. A straightforward way to visualize the structure is to use a hexagonal lattice with the unit cell being a quintuple layer, as shown in Fig. \ref{structure}(a). Like graphite, adjacent Se-Se (Te-Te) layers are hold together by weak van der Waals force.

The XRD patterns were recorded using a Huber G670 imaging-plate Guinier camera equipped with a Ge monochromator and Cu K$_{\alpha1}$ radiation (1.54059 \text{\AA}). Data were collected in steps of 0.005$^\circ$ in a wide temperature range from 10 K to 270 K. The lattice parameters were calculated via WINPREP program with residual factor $0.02 < R_w < 0.03$. Typical XRD spectra of Bi$_2$Se$_3$ and Sb$_2$Te$_3$ crystals (at 10 K) are shown in Fig.\ref{structure}(b) and (c), respectively. From the spectra, one can determine the lattice constant of Bi$_2$Se$_3$ (extrapolated to 0 K): $a_{\text{hex}}=4.1263$ \text{\AA} and $c_{\text{hex}}=28.481$ \text{\AA}, which translate to $a_{\text{rho}}=9.7880$ \text{\AA} and $\alpha_{\text{rho}}=24.337^\circ$ for the rhombohedral cell. For Sb$_2$Te$_3$, the corresponding values are $a_{\text{hex}}=4.2423$ \text{\AA}, $c_{\text{hex}}=30.191$ \text{\AA}, and $a_{\text{rho}}=10.357$ \text{\AA}, $\alpha_{\text{rho}}=23.635^\circ$.
\begin{figure}
\includegraphics{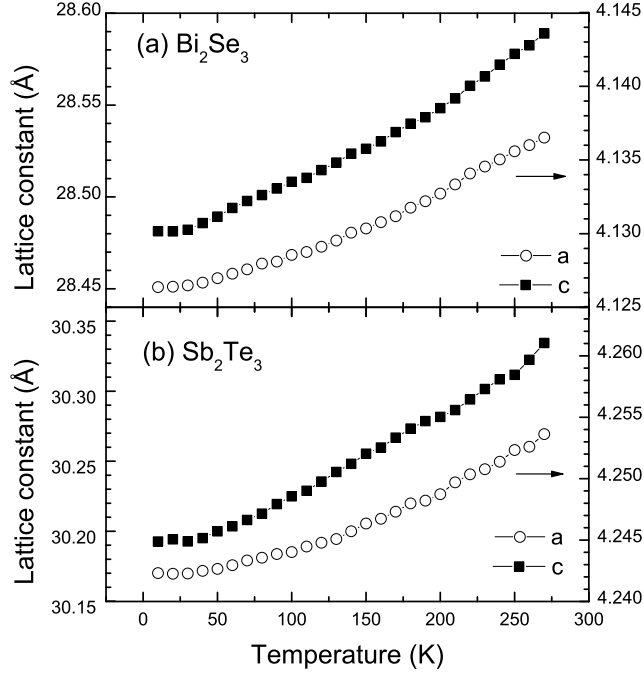}
\caption{\label{lattice} Lattice constant of (a) Bi$_2$Se$_3$ and (b) Sb$_2$Te$_3$ as a function of temperature.}
\end{figure}

\begin{figure}
\includegraphics{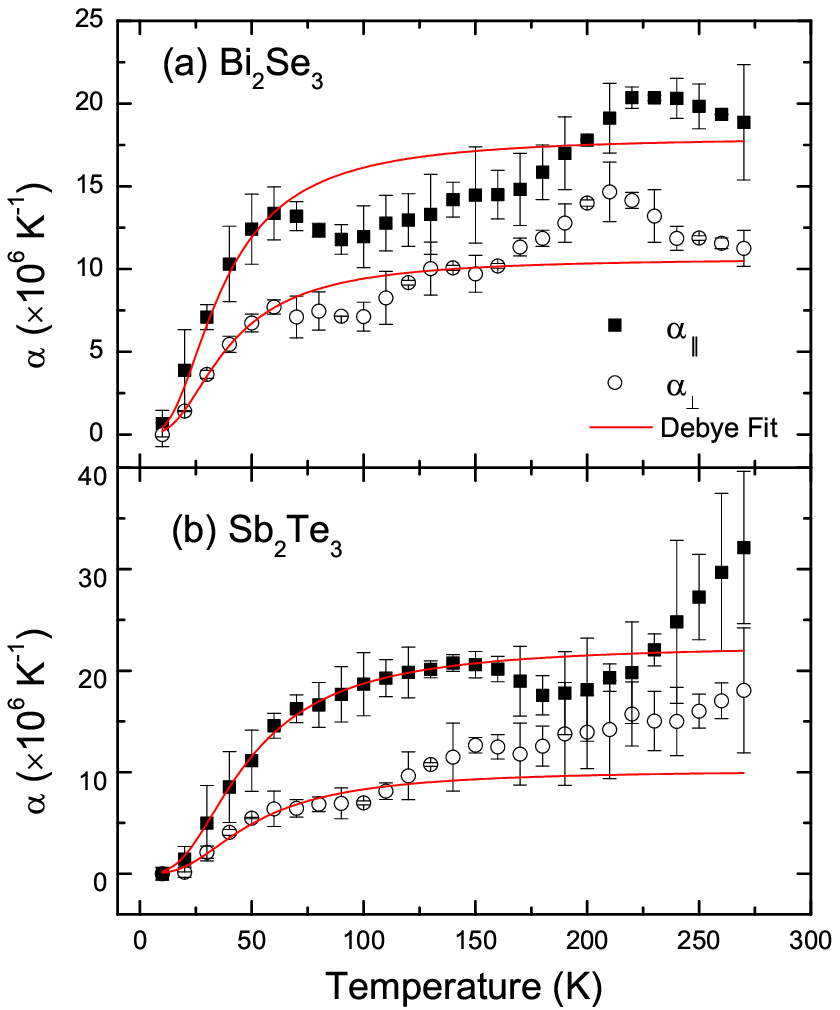}
\caption{\label{thermal}(color online) Linear thermal expansion coefficients of (a) Bi$_2$Se$_3$ and (b) Sb$_2$Te$_3$ as a function of temperature. Solid lines represent best fits using Debye model. The data well agree with the Debye $T^3$ law at low temperatures, while considerable deviation is evidenced at above 150 K.}
\end{figure}

Figure \ref{lattice} plots the measured lattice parameters of Bi$_2$Se$_3$ and Sb$_2$Te$_3$ as a function of temperature, from which the linear thermal expansion coefficients ($\alpha$) of the material can be derived
\begin{equation}
\alpha_L=\frac{1}{L}\frac{dL}{dT},
\end{equation}
where $L$ is the lattice constant (i.e., $a_{\text{hex}}$ or $c_{\text{hex}}$), and $T$ is the temperature. Figure \ref{thermal} shows the obtained linear thermal expansion coefficients of Bi$_2$Se$_3$ and Sb$_2$Te$_3$ parallel ($\alpha_\parallel$) and perpendicular ($\alpha_\perp$) to the trigonal $c$-axis. Considerable anisotropy between $\alpha_\parallel$ and $\alpha_\perp$ is observed, reflective of the material bonding anisotropy. The ratio of the elastic constants $C_{11}/C_{13}$ characterizes the anisotropy of chemical bonds; $C_{11}/C_{13}=2.5$ for Bi$_2$Te$_3$ \cite{semihandbook} and similar values are expected for Bi$_2$Se$_3$ and Sb$_2$Te$_3$ owing to the structural similarity and comparable Debye temperature \cite{bite}. To understand the temperature dependence of $\alpha(T)$, it is instructive to introduce the Gr\"uneisen relation\cite{gru}
\begin{equation}
c_v\gamma=\kappa \beta,    \label{gruneisen}
\end{equation}
where $c_v$ is the specific heat, $\gamma$ is the Gr\"uneisen parameter, $\kappa$ is the bulk modulus, and $\beta=2\alpha_\parallel+\alpha_\perp$ is the volumetric thermal expansion coefficient. Usually, $\gamma$ and $\kappa$ are only weakly dependent on the temperature \cite{SolidState}, indicating that the thermal expansion coefficient has the same temperature dependence as the specific heat. This relationship ($\alpha(T)$ vs. $c_v(T)$) remains valid for anisotropic materials \cite{gru2}, therefore we can apply Debye model to fit our data
\begin{equation}
\alpha=\alpha_0(\frac{T}{\Theta_D})^3\int_0^{\Theta_D/T}\frac{x^4e^xdx}{(e^x-1)^2},    \label{gruneisen2}
\end{equation}
where $\alpha_0$ is a temperature independent fitting parameter, and $\Theta_D$ is the Debye temperature. Good agreement is achieved in the low temperature regime, as shown by the solid lines in Fig. \ref{thermal}, consistent with that observed in Bi$_2$Te$_3$ \cite{Barnes_pla}. The obtained Debye temperature (from fitting both $\alpha_\parallel$ and $\alpha_\perp$) is $\Theta_D=160$ K for Bi$_2$Se$_3$ and $\Theta_D=200$ K for Sb$_2$Te$_3$, close to the literature values (182 K and 160 K, respectively) \cite{semihandbook}. At above 150 K, the experimental results deviate from the Debye model. Similar anomalous behavior is also evident in Bi$_2$Te$_3$ \cite{britjap}, but the origin is not conclusive \cite{Barnes_pla}. There are two possible explanations: (1) This is due to higher-order anharmonic effects. Generally, the anharmonicity of the bonding forces in a crystal is characterized by the Gr\"uneisen parameter. Higher-order anharmonic effects may lead to a nontrivial temperature dependence $\gamma(T)$, giving rise to anomalous $\alpha_\parallel(T)$ and $\alpha_\perp(T)$ through the Gr\"uneisen relation. This scenario is used to explain the thermal expansion of tellurium \cite{Te,gibbons} and Bi$_2$Te$_3$ \cite{pavlova}. (2) The deviation could be due to the breaking of the van der Waals bond between two Se-Se (Te-Te) layers, as proposed in Ref.[\onlinecite{pavlova}] and [\onlinecite{britjap}] for Bi$_2$Te$_3$. The van der Waals binding energy is typically in the range of 0.4-4 kJ/mole, while the energy of thermal motion at room temperature is about 2.5 kJ/mole, suggesting that the degradation of the van der Waals bond may cause the anomaly in $\alpha(T)$ at levitated temperatures. These two reasons may also explain the differences in $\alpha(T)$ between Bi$_2$Se$_3$ and Sb$_2$Te$_3$, but further experimental and theoretical work is needed to fully understand it.

Finally, we summarize the room temperature thermal expansion parameters of Bi$_2$Se$_3$, Sb$_2$Te$_3$ and Bi$_2$Te$_3$ in Table \ref{table}, as a reference for future study of TIs and device engineering. The thermal expansion coefficients for Bi$_2$Se$_3$ and Sb$_2$Te$_3$ are taken from Fig.\ref{thermal}, which are in good agreement with the values documented in the materials handbook \cite{semihandbook}. Gr\"uneisen parameters are calculated from Eq.[\ref{gruneisen}]. Specific heat at constant pressure ($c_p$) of the materials is better documented in the literature, and thus used in the calculation \cite{remark}. For Bi$_2$Te$_3$ the values are extracted from various previous works.

\begin{table*}
\caption{\label{table}Thermal expansion coefficients and Gr\"uneisen parameters of Bi$_2$Se$_3$ and Sb$_2$Te$_3$ (at 270 K), and Bi$_2$Te$_3$ (at room temperature).}
\begin{ruledtabular}
\begin{tabular}{lcccccc}
&$\alpha_\parallel$ 		& $\alpha_\perp$			 &  $\beta$				 &   $c_p$		& $\kappa$	 & $\gamma$\\
&$(\times10^{-5}$ K$^{-1})$	& $(\times10^{-5}$ K$^{-1})$ 	&  $(\times10^{-5}$ K$^{-1})$ 	&  (JK$^{-1}$mol$^{-1}$) 	& (GPa)	& \\
\hline
Bi$_2$Se$_3$ 	& $1.9\pm0.3$			 & $1.1\pm0.1$			 & $4.1$	& $124.3$\footnote{See Ref.[\onlinecite{semihandbook}]}	 & $48.4$\footnote{See Ref.[\onlinecite{springermat}]}	 & $1.4$\\
Sb$_2$Te$_3$ 	 & $3.2\pm0.7$			 & $1.8\pm0.6$			 & $6.8$	& $128.8^\text{a}$	 & $44.8$\footnote{See Ref.[\onlinecite{kappa}]}		 & $2.3$\\
Bi$_2$Te$_3$	 & $2.1^\text{a}$			 & $1.4^\text{a}$	 		& $4.9$ 	& $124.4$\footnote{See Ref.[\onlinecite{mills}]}		 & $37.4$\footnote{See Ref.[\onlinecite{bite}]}		 & $1.5^\text{e}$\\
\end{tabular}
\end{ruledtabular}
\end{table*}

In conclusion, we have measured the thermal expansion parameters of Bi$_2$Se$_3$ and Sb$_2$Te$_3$ crystals in a wide temperature range from 10 K to 270 K. The extracted linear thermal expansion coefficients are found in consistent with the Debye model at low temperatures, but deviate at above 150 K. Our result is crucial for interpreting the temperature dependent Raman shift in Bi$_2$Se$_3$ and Sb$_2$Te$_3$ \cite{Raman_Kim}, and may provide insight for understanding the thermal properties of TIs.

We would like to thank S. Zhou for helping with the chemical structure in Fig. 1(a). This work is supported by the DOE (DE-FG02-07ER46451). The XRD measurement was performed at the National High Magnetic Field Laboratory, which is supported by NSF Cooperative Agreement No. DMR-0654118, by the State of Florida, and by the DOE.


\begin{thebibliography}{text}
\bibitem{remark}The difference between $c_p$ and $c_v$ is negligible \cite{springermat}, compared with the uncertainty of our measurements.
\end{thebibliography}
\end{document}